\documentclass[11pt]{article}
\usepackage{amsmath}
\usepackage{amsthm}
\usepackage{graphicx}
\usepackage{amssymb}

\usepackage{multicol}
\usepackage{wrapfig}
\usepackage{paralist} 
\usepackage{graphics} 
\usepackage{epsfig} 
\usepackage[dvips]{color}
\usepackage[sc,small]{caption}

\definecolor{RED}{rgb}{1,0,0}
\definecolor{BLUE}{rgb}{0,0,1}
\definecolor{PURPLE}{rgb}{1,0,1}

\DeclareMathOperator{\Lk}{{}_{L}}
\DeclareMathOperator{\CaH}{{}_{CaH}}
\DeclareMathOperator{\NaF}{{}_{NaF}}
\DeclareMathOperator{\KDR}{{}_{KDR}}
\DeclareMathOperator{\KM}{{}_{KM}}
\DeclareMathOperator{\nA}{\, {nA}}

\newcommand{\eps}{\varepsilon}
\newcommand{\sech}{\ensuremath{\mathrm{sech\ }}}
\newcommand{\gCa}{\ensuremath{g_\mathrm{Ca}}}

\begin{document}
\author{John~Burke${}^{\dag}$\footnote{Corresponding author: { jb@math.bu.edu}}, Mathieu Desroches${}^{\ddag}$, Anna~M.~Barry${}^{\dag}$, Tasso~J.~Kaper${}^{\dag}$, and Mark~A.~Kramer${}^{\dag}$ \\
${}^{\dag}$\small Department of Mathematics and Statistics, Center for BioDynamics, Boston University, Boston, MA 02215, USA \\ 
${}^{\ddag}$\small Department of Engineering Mathematics, University of Bristol, Bristol, UK
}
\title{A showcase of torus canards in neuronal bursters}
\date{\today}
\maketitle

\begin{abstract}

Rapid action potential generation --- spiking --- and alternating intervals of spiking and quiescence --- bursting --- are two dynamic patterns observed in neuronal activity.  In computational models of neuronal systems, the transition from spiking to bursting often exhibits complex bifurcation structure.  One type of transition involves the torus canard, which was originally observed in a simple biophysical model of a Purkinje cell. In this article, we expand on that original result by showing that torus canards arise in a broad array of well-known computational neuronal models with three different classes of bursting dynamics:  sub-Hopf/fold cycle bursting, circle/fold cycle bursting, and fold/fold cycle bursting. The essential features that these models share are multiple time scales leading naturally to decomposition into slow and fast systems, a saddle-node of periodic orbits in the fast system, and a torus bifurcation in the full system.  We show that the transition from spiking to bursting in each model system is given by an explosion of torus canards.  Based on these examples, as well as on emerging theory, we propose that torus canards are a common dynamic phenomenon separating the regimes of spiking and bursting activity.

\end{abstract}


\section{Introduction}

The primary unit of brain electrical activity --- the neuron --- generates a characteristic dynamic behavior: when excited sufficiently, a rapid (on the order of milliseconds) increase then decrease in the neuronal voltage occurs, see for example Ref.~\cite{HH}.  This action potential (or `spike') mediates communication between neurons, and therefore is fundamental to understanding brain activity~\cite{Dayan,Koch,Rieke}.  Neurons exhibit many different types of spiking behavior including regular periodic spiking and bursting, which consists of a periodic alternation between intervals of rapid spiking and quiescence, or active and inactive phases, respectively~\cite{Conners,coombes,Izhikevich}. Bursting activity may serve important roles in neuronal communication, including robust transmission of signals and support for synaptic plasticity~\cite{Izhikevich2003,Lisman}.

Computational models of spiking and bursting allow a detailed understanding of neuronal activity.  Perhaps the most famous computational model in neuroscience --- developed by Hodgkin and Huxley in 1952~\cite{HH} --- provided new insights into the biophysical mechanisms of spike generation.  Only later were the dynamical processes that support spiking and bursting explored, see for example Refs.~\cite{IzhikevichBook,Rinzel,RE}.  Recent research has led to a number of classification schemes of bursting, including a scheme by Izhikevich~\cite{Izhikevich} based on the bifurcations that support the onset and termination of the burst's active phase.  This classification requires identifying the separate time scales of the bursting activity:  a fast time scale supporting rapid spike generation, and a slow time scale determining the duration of the active and inactive burst phases. This separation of time scales naturally decomposes the full model into a fast system and a slow system. Understanding the bifurcation structure of the isolated fast system is the principal element of the classification scheme.  Typically, the onset of the burst's active phase corresponds to a loss of fixed point stability in the fast system, and the termination of the active phase to a loss of limit cycle stability in the fast system.  For example, in a fold/fold cycle burster, the former transition occurs through a saddle-node bifurcation (or fold) of attracting and repelling fixed points in the fast system, and the latter transition occurs through a fold of attracting and repelling limit cycles in the fast system. We shall refer to this classification scheme for most of the bursters discussed here. Also, we refer the reader to Ref.~\cite{GJK2001} for a natural catalog of the bifurcations that can initiate and terminate bursting in fast-slow systems. There, low-codimension singularities in the fast systems are analyzed in a systematic fashion, and the slow variables are used as the unfolding parameters. The natural catalog is generated by identifying all possible paths that lead to bursting in these unfolding spaces.

Although the dynamics of spiking and bursting have been studied in detail, the mathematical mechanisms that govern transitions between neuronal states are only now beginning to be understood. The transition from spiking to bursting activity has been shown to involve different mechanisms including the blue sky catastrophe~\cite{Shilnikov}, period doubling~\cite{CymbalyukShilnikov}, chaos~\cite{Medvedev,Terman2}, and mixed-mode oscillations~\cite{Wojcik}.  Recently, it has been proposed that the transition from spiking to bursting can also involve torus canards~\cite{Benes,Kramer}.  In these models, limit cycles in the fast system terminate in a fold. However, these models exhibit unexpected behavior:  the dynamics of the full system pass through the fold of limit cycles, but the burst's active phase does not terminate.  Instead, the dynamics of the full system move through the fold of limit cycles and follow the branch of repelling limit cycles for some time, resulting in a torus canard. 

In this article, we demonstrate that torus canards arise naturally in computational neuronal models of multiple time scale type. In particular, we show that they arise in well-known neuronal models exhibiting three different classes of bursting: sub-Hopf/fold cycle bursting, circle/fold cycle bursting, and fold/fold cycle bursting. These models are all third order dynamical systems with two fast and one slow variable. We show that these models all have saddle-node bifurcations of periodic orbits (a.k.a. folds of limit cycles) in the fast systems, and torus bifurcations in the full systems. In addition, we show that the transitions from spiking to bursting in these systems are given by explosions of torus canards, as well as by some related mixed-mode oscillations (MMO). Based on these observations, we propose that torus canard explosions are a commonly-occurring transition mechanism from spiking to bursting in neuronal models.

The organization of this manuscript is as follows. In Section~\ref{sec:overview}, we review the torus canard phenomenon identified in Ref.~\cite{Kramer} and recently studied in Ref.~\cite{Benes}. In Sections~\ref{sec:HR}--\ref{sec:WCI}, we present the main results about torus canards in the transitions from spiking to bursting in three well-known neuronal models. Finally, our conclusions are presented in Section~\ref{sec:conclusions}.

{\bf Remark}: Throughout this article, we make extensive use of the software package AUTO~\cite{AUTO} to carry out the continuation of fixed points and periodic orbits of the models and their fast systems. Bursting trajectories are found using direct numerical simulations with a stiff-solver suited to multiple time scale systems, starting from arbitrary initial conditions, and we disregard transients in the figures.

\section{Overview of Torus Canards}\label{sec:overview}

In this section, we briefly review the classical phenomenon of canards
as they arise, for example, in the van der Pol oscillator, and the
recently-identified phenomenon of torus canards as they arise in a Purkinje cell model.

The van der Pol oscillator, in the relaxation limit, is the paradigm
example of a system with a canard.  The system consists of one fast
variable and one slow variable, and one adjustable parameter~\cite{Diener, Eckhaus}.  For most parameter values, the oscillator's dynamics exhibit either a fixed point or a relaxation oscillation, in which the dynamics of the full system alternate between two branches of attracting fixed points in the fast system.  Canard orbits exist over a small parameter range in the transition regime between these two extremes. The oscillations are born in a supercritical Hopf bifurcation of the full system which yields small amplitude oscillations near onset. The first canard orbits (referred to as `headless ducks') occur when
the dynamics of the full system pass through a fold of fixed points of the fast system (where branches of attracting and repelling fixed points meet) and follow the branch of repelling fixed points for some time before returning to the  attracting branch.  With further changes in the bifurcation parameter, the oscillations grow in amplitude and move further along the repelling branch, eventually reaching a second fold of fixed points of the fast system (corresponding to the maximal canard).  Beyond this parameter value, the dynamics leave the branch of repelling fixed points and transition to the other attracting branch of fixed points (forming a `duck with a head').  As the parameter increases further, the dynamics leave the repelling branch sooner, eventually resulting in a relaxation oscillation.  These changes in the oscillation amplitude --- from small amplitude oscillation to `headless duck', `duck with head', and finally large amplitude relaxation oscillations --- occur over a small parameter range and are labelled a canard explosion~\cite{Diener, Dumortier, Eckhaus, Krupa}.

In the classical canard described above, the dynamics of the full system undergo a Hopf bifurcation and, after passing through a fold of fixed points in the fast system, follow a branch of repelling fixed points for some time.  The torus canard is the one-dimension-higher analog of this classical canard. In a torus canard, the dynamics of the full system undergo a torus bifurcation (instead of a Hopf bifurcation) and, after passing through a saddle-node bifurcation of periodic orbits (instead of fixed points) in the fast system, follow a branch of repelling periodic orbits for some time.  We now briefly review the essential features of the torus canards in the Purkinje cell model~\cite{Kramer}. This single-compartment model consists of five ordinary differential equations that describe the dynamics of the membrane potential, $V$, and four ionic gating variables, $m_{\CaH}$, $h_{\NaF}$, $m_{\KDR}$, and $m_{\KM}$:
\begin{subequations} \label{eq:Purkinje}
\begin{align}
 C \, \dot{V} &= -J - g_{\Lk} (V-V_{\Lk}) - g_{\CaH} m_{\CaH}^2 (V-V_{\CaH}) - g_{\NaF} m_{\NaF,\infty}^3 h_{\NaF} (V-V_{\NaF}) \label{eq:diffeqV} \, , \\
 & \qquad {} - g_{\KDR} m_{\KDR}^4 (V-V_{\KDR}) - g_{\KM} m_{\KM} (V-V_{\KM}) \notag \, , \\
 \dot{m}_{\CaH} &= \alpha_{\CaH} (1-m_{\CaH}) - \beta_{\CaH} m_{\CaH} \label{eq:diffeqCaH} \, , \\
 \dot{h}_{\NaF} &= \alpha_{\NaF} (1-h_{\NaF}) - \beta_{\NaF} h_{\NaF} \label{eq:diffeqNaF} \, , \\
 \dot{m}_{\KDR} &= \alpha_{\KDR} (1-m_{\KDR}) - \beta_{\KDR} m_{\KDR} \label{eq:diffeqKDR} \, , \\
 \dot{m}_{\KM}  &= \alpha_{\KM}  (1-m_{\KM} ) - \beta_{\KM}  m_{\KM} \label{eq:diffeqKM} \, . 
\end{align}
\end{subequations}

The parameter $J$ represents an externally applied current. The forward and backward rate functions ($\alpha_X$ and $\beta_X$ for $X={\rm CaH, NaF, KDR, KM}$) and fixed parameter values are given in the appendix of Ref.~\cite{Kramer}. The gating variable $m_{\KM}$ for the muscarinic receptor suppressed potassium current (a.k.a. $M$ current) evolves on a much slower time scale than the other variables. Hence, the dynamics of system~(\ref{eq:Purkinje}) may be studied using the four-dimensional fast system, which is defined by setting ${\dot m}_{\KM}=0$ and treating $m_{\KM}$ as a bifurcation parameter.

\begin{figure}[t!]
\begin{center}
\resizebox{6.2in}{!}{\includegraphics{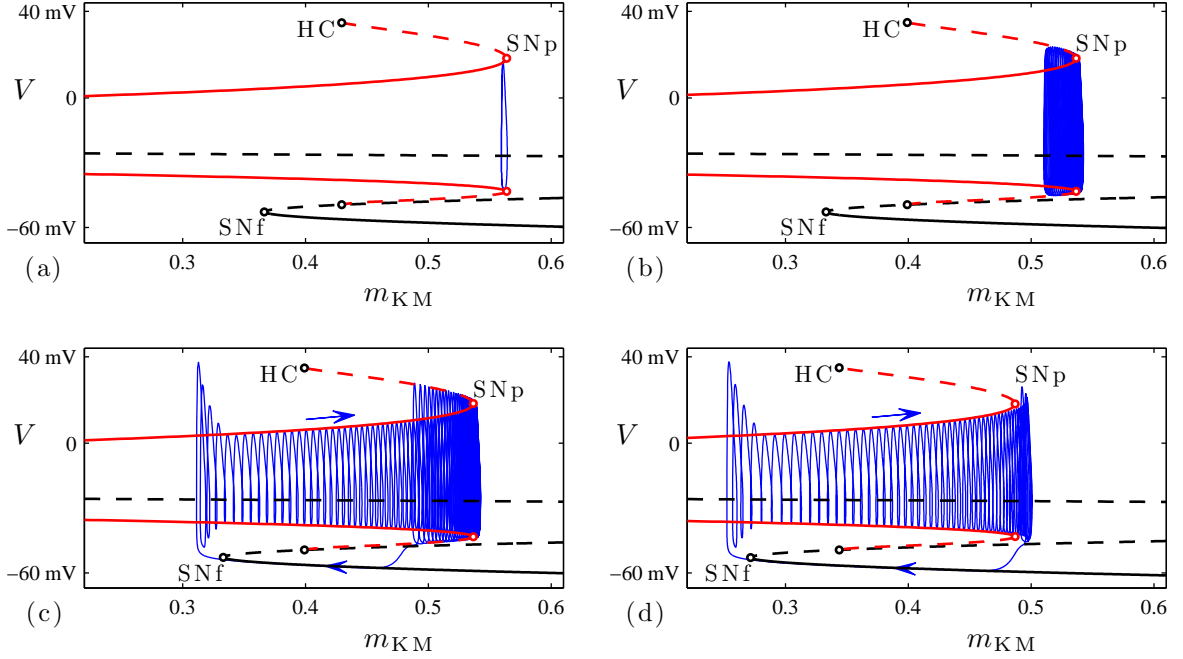}}
\end{center}
\caption{Dynamics of the Purkinje cell model~(\ref{eq:Purkinje}) at several values of $J$: (a) rapid spiking, at $J=-34\nA$; (b) amplitude modulated spiking or `headless duck' torus canard, at $J=-32.94\nA$; (c) `duck with head' torus canard at $J = -32.93815\nA$; (d) standard fold/fold cycle bursting at $J = -31\nA$. In each frame, the trajectory of the full system (blue curve) is plotted in projection in the $(m_{\KM}, V)$ phase space, along with the bifurcation diagram of the fast system at the corresponding value of $J$. The bifurcation diagrams include branches of fixed points (black curves) and periodic orbits (two red curves, indicating maximal and minimal values of $V$ over the orbit). Solid/dashed curves indicate stable/unstable solutions of the fast system. The labels mark saddle-node bifurcations of fixed points (SNf), saddle-node bifurcations of periodic orbits (SNp), and homoclinic bifurcations (HC). Arrows indicate the direction of drift in $m_{\KM}$ for the trajectories.}
\label{fig:Purkinje}
\end{figure}

Figure~\ref{fig:Purkinje} illustrates the transition from spiking to bursting through the torus canard explosion in system~(\ref{eq:Purkinje}) as parameter $J$ increases.  For $J$ sufficiently negative, the full system exhibits rapid spiking,  as shown in Fig.~\ref{fig:Purkinje}a superimposed on the corresponding bifurcation diagram of the fast system.  The spiking orbit in the full system (blue) remains near the branch of attracting periodic orbits (solid red) in the fast system.  At larger values of $J$, the torus canard orbit first emerges as a transition to amplitude modulated (AM) spiking in the full system (Fig.~\ref{fig:Purkinje}b).  During AM spiking, the full system orbit oscillates near the branch of attracting periodic orbits of the fast system and the slow variable $m_{\KM}$ increases.  The full system dynamics reach the saddle-node (or fold) of periodic orbits (SNp) --- in which the attracting and repelling branches of period orbits meet --- and continue near the branch of repelling periodic orbits as $m_{\KM}$ decreases.  Eventually, the full system dynamics return to the neighborhood of the attracting periodic orbits, restarting the AM sequence.  The full system~(\ref{eq:Purkinje}) possesses a torus bifurcation at this transition, which marks the onset of the regime of torus canards. This is the analog in one higher dimension of the Hopf bifurcation which marks the onset of classical canards.

Further increases in $J$ produce a transition from AM spiking to bursting orbits, in which the full system dynamics leave the branch of repelling periodic orbits for the branch of attracting fixed points (black curve) in the fast system (Fig.~\ref{fig:Purkinje}c).  This passage to the branch of attracting fixed points corresponds to the onset of the inactive burst phase, during which the slow variable $m_{\KM}$ decreases.  Eventually, the full system reaches the saddle-node of fixed points in the fast system (SNf) and transitions to the branch of attracting periodic orbits in the fast system to begin the active phase of the burst.  The transition from AM spiking to bursting corresponds to the progression from `headless ducks' (AM spiking) to `ducks with heads' (bursting) in the torus canard sequence.  During this transition mixed-mode oscillations (MMO) appear, which consist of repeating sequences of AM spiking and bursting orbits.  Finally, at large enough values of $J$, the full system dynamics exhibit the standard fold/fold cycle bursting~\cite{Izhikevich}, in which the active phase of the burst begins at a saddle-node of fixed points SNf and ends at saddle-node of periodic orbits SNp (Fig.~\ref{fig:Purkinje}d).

The existence of torus canard-like trajectories was described in Ref.~\cite{IzhikevichSIAM2000} in the context of an abstract model, consisting of a planar fast-slow system that is rotated about an axis. The behavior of torus canards in a similar abstract model that breaks the rotational symmetry was considered in Ref.~\cite{Benes}.
Just as in the case of the Purkinje cell model of Ref.~\cite{Kramer} described above, the key ingredients in these abstract models are a fold of limit cycles in the fast system and a torus bifurcation in the full system. Moreover, the torus canards in the abstract model of Ref.~\cite{Benes} also undergo an explosion involving headless ducks, MMO, and ducks with heads, and they occur in the transition regime between spiking and bursting.

\section{Torus Canards in the Hindmarsh-Rose System} \label{sec:HR}

We begin with the following modified version of the Hindmarsh-Rose (HR) system~\cite{HindmarshRose} developed in Ref.~\cite{Tsaneva} 
\begin{subequations} \label{eq:HR}
\begin{eqnarray}
\dot{x} &=& s a x^3 - s x^2 - y - b z,\\
\dot{y} &=& \phi(x^2-y),\\
\dot{z} &=& \eps( s a_1 x + b_1 - k z ).
\end{eqnarray}
\end{subequations}

The small parameter $\eps$ induces a separation of time scales, so that the
voltage variable $x$ and the gating variable $y$ are fast and the recovery
variable $z$ is slow.

The HR model is known to exhibit rich dynamics, including square-wave bursting (a.k.a. plateau bursting) and pseudo-plateau bursting~\cite{Tsaneva}. Here, we show that this model also exhibits sub-Hopf/fold cycle bursting (in which the active phase of the burst initiates in a subcritical Hopf bifurcation and terminates in a fold of limit cycles), and that torus canards occur precisely in the transition region from spiking to this type of bursting.

In this section, we first identify the parameter regimes in which the fast system of the HR model has a saddle-node of periodic orbits (Section~\ref{sec:HR_fast}) and in which the full HR model has a torus bifurcation (Section~\ref{sec:HR_tr}). Once these key ingredients are identified, we show (Section~\ref{sec:HR_tc}) that the full HR model includes a torus canard explosion, and that it lies in the transition region between spiking and bursting.

We treat $b_1$ as the primary control parameter, meaning that we examine the transition from spiking to bursting as $b_1$ varies. We take $s$ as a secondary control parameter, and examine how the transition from spiking to bursting behaves at different values of $s$. Except where otherwise noted, we set the remaining parameters to
\begin{gather} \label{eq:HR_params}
 a=0.5\, , \quad \phi=1\, , \quad a_1=-0.1\, , \quad k=0.2 \, , \quad b=10 \, , \quad \eps = 10^{-5} \, ,
\end{gather}
which is based on the values used in Ref.~\cite{Tsaneva}.

\subsection{Bifurcation Analysis of the Fast System} \label{sec:HR_fast}

\begin{figure}[t!]
\begin{center}
\resizebox{6.2in}{!}{\includegraphics{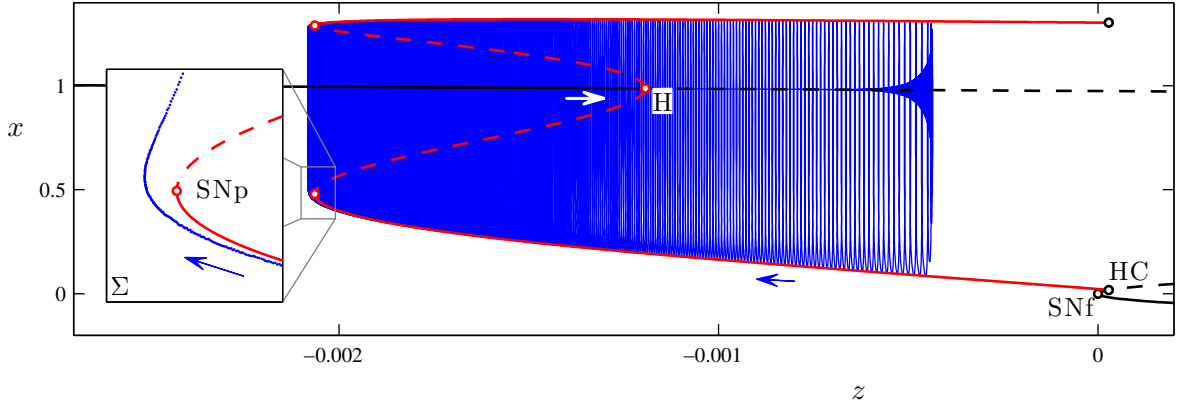}}
\end{center}
\caption{An example of sub-Hopf/fold cycle bursting in the HR system~(\ref{eq:HR}), with $(b_1, s)=(-0.162,-1.95)$. The other parameter values are given by~(\ref{eq:HR_params}). The bursting trajectory (blue curve) is plotted in projection onto the $(z,x)$ phase space, along with the bifurcation diagram of the fast system at this value of $s$. The bifurcation diagram includes branches of fixed points and periodic orbits, and follows the plotting conventions in Fig.~\ref{fig:Purkinje}. The inset shows the Poincar\'{e} map of the bursting trajectory near SNp, also plotted in projection onto the $(z,x)$ phase space. The Poincar\'{e} surface $\Sigma \equiv \{(x,y,z) | 0=s a x^3-s x^2 - y - bz\}$ is chosen so that the iterates correspond to local extrema in $x$ of the trajectory.}
\label{fig:HR_SHFCburst}
\end{figure}

The fast system of~(\ref{eq:HR}) is obtained by setting $\eps=0$. It is independent of $b_1$, so at fixed $s$ the slow variable $z$ serves as the bifurcation parameter. Figure~\ref{fig:HR_SHFCburst} shows a bifurcation diagram of the fast system at fixed $s=-1.95$. The branch of fixed points is stable for large negative $z$ values. As $z$ increases, the fixed point loses stability in a subcritical Hopf bifurcation (H), undergoes a saddle-node bifurcation at large $z$ (not shown in the figure), then regains stability in a second saddle-node bifurcation (SNf). Fixed points between the two saddle-node bifurcations are of saddle-type, with one stable and one unstable eigenvalue. The branch of repelling periodic orbits created in the Hopf bifurcation undergoes a saddle-node bifurcation (SNp) then terminates in a homoclinic bifurcation (HC) --- i.e., a homoclinic connection to the saddle fixed point.

Figure~\ref{fig:HR_SHFCburst} also includes a trajectory of the full HR system which illustrates sub-Hopf/fold cycle bursting. The trajectory is plotted in projection onto the $(z,x)$ phase space and superimposed on the bifurcation diagram of the fast system. During the quiescent phase of the burst, the trajectory of the full system drifts up in $z$ along the branch of fixed points of the fast system. The active phase of the burst initiates when the trajectory passes through the subcritical Hopf bifurcation H and, after a slow passage effect~\cite{N87,N88} (which causes the orbit to stay near the branch of repelling fixed points for some time), spirals out to the attracting branch of periodic orbits. During the active phase of the burst, the trajectory shadows the attracting branch of periodic orbits as it drifts to smaller $z$ values. The active phase terminates when the trajectory falls off the branch of periodic orbits at SNp and spirals back in toward the attracting branch of fixed points to repeat the cycle.

\begin{figure}[t!]
\begin{center}
 \resizebox{4.5in}{!}{\includegraphics{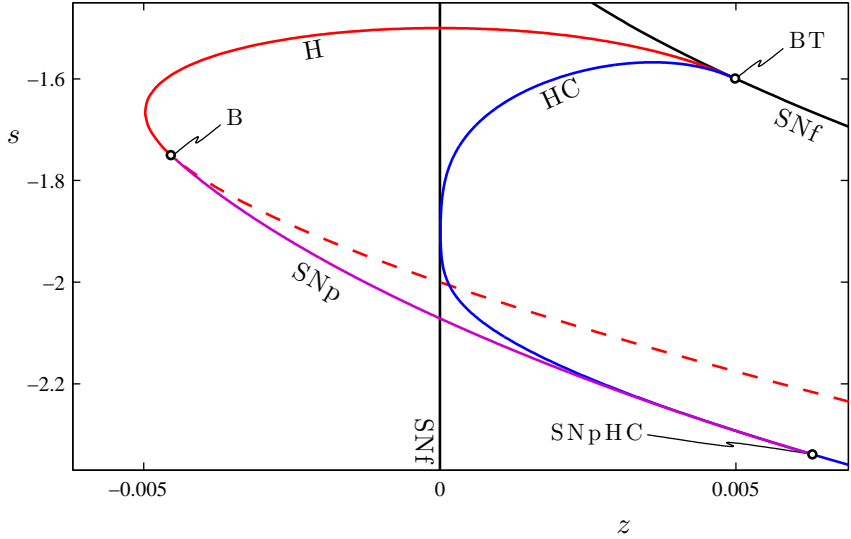}}
\end{center}
\caption{Two-parameter bifurcation diagram of the fast system of~(\ref{eq:HR}) in the $(z,s)$-plane. The loci of Hopf H (red curve) and homoclinic HC (blue curve) bifurcations emerge from the Bogdanov-Takens point BT at $(z,s) \simeq (0.004985, -1.599)$. The curve H is plotted as a solid/dashed line when the Hopf bifurcation is supercritical/subcritical. The saddle-node of periodic orbits SNp (purple curve) exists between the Bautin bifurcation point B at $(z,s) \simeq (-0.004541, -1.75)$ and the SNpHC at $(z,s) \simeq (0.006291, -2.339)$.}
\label{fig:HR_twopar2d}
\end{figure}

With the default choice of parameters, the HR system already exhibits a key feature required for torus canards: a saddle-node of periodic orbits in the fast system. To further explore the range over which torus canards may occur, we consider how the bifurcation structure of the fast system, as shown in Fig.~\ref{fig:HR_SHFCburst}, changes with the parameter $s$. To this end, we compute loci of the codimension-1 bifurcations H, SNf, SNp and HC, in the $(z,s)$ parameter plane of the fast system. The results are shown in Fig.~\ref{fig:HR_twopar2d}. There are three noteworthy codimension-2 bifurcation points included in this figure. The loci of Hopf and homoclinic bifurcations emerge from the saddle-node of fixed points at a Bogdanov-Takens point (BT). A Bautin bifurcation (B) marks the point at which the Hopf bifurcation changes from supercritical to subcritical, and also the associated emergence of the curve of saddle-node of periodic orbits. Finally, this SNp curve ends when it collides with the homoclinic bifurcation at the point labeled SNpHC. This final codimension-2 bifurcation amounts to a change in the criticality of the homoclinic bifurcation. Thus, the HR system includes a saddle-node of periodic orbits for values of $s$ within the range $-2.3388 \leq s \leq -1.75$, and it is within this range that we expect the system may also include torus canards.

\subsection{Torus Bifurcation in the Full System} \label{sec:HR_tr}

\begin{figure}[t!]
 \center
 \resizebox{3.2in}{!}{\includegraphics{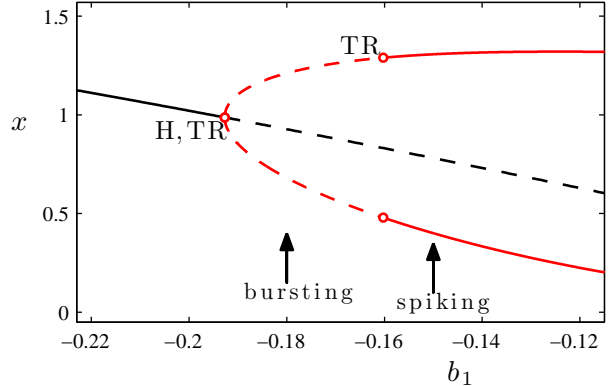}}
  \caption{Bifurcation diagram of the HR system~(\ref{eq:HR}) at $s=-1.95$, including branches of fixed points (black curve) and periodic orbits (two red curves, indicating maximal and minimal values of $x$ over the orbit). Solid/dashed curves indicate stable/unstable solutions. The torus bifurcation at $b_1 \simeq -0.1603$ is supercritical, leading to bursting for smaller values of $b_1$.}
  \label{fig:HR_bif3d}
\end{figure}

The second key ingredient to the emergence of torus canards is the presence of a torus bifurcation in the full system, between the regimes of rapid spiking and bursting. To see that this does occur in the HR system~(\ref{eq:HR}), consider the bifurcation diagram of the full system shown in Fig.~\ref{fig:HR_bif3d} at fixed $s=-1.95$. As $b_1$ increases, the branch of stable fixed points undergoes a supercritical Hopf bifurcation at $b_1 \simeq -0.1927$, creating a branch of stable periodic orbits. This branch of periodic orbits changes stability in two torus bifurcations, the first of which occurs near the Hopf bifurcation where the periodic orbits are of very small amplitude. Beyond the second torus bifurcation at $b_1 \simeq -0.1603$, the periodic orbits are stable and correspond to the rapid spiking state of the system. It is this upper torus bifurcation, which lies between the regimes of spiking and bursting, that is associated with torus canards. Moreover, continuation in the secondary parameter $s$ shows that this upper torus bifurcation persists over the entire range of $s$ values for which the fast system exhibits a saddle-node of periodic orbits (i.e., $-2.339 \leq s \leq -1.75$). We therefore expect the torus canard phenomenon to occur in a neighborhood of SNp over this entire range of $s$ values.

\subsection{Torus Canard Explosion} \label{sec:HR_tc}

The transition from spiking to bursting as $b_1$ decreases through the torus bifurcation at $b_1 \simeq -0.1603$ occurs by way of a torus canard explosion. When $b_1$ is above the torus bifurcation, the periodic orbit of the full system is stable. This trajectory resembles a periodic orbit taken from the attracting branch of periodic orbits of the fast system at a value of $z$ near SNp (refer to the bifurcation diagram in Fig.~\ref{fig:HR_SHFCburst}). As $b_1$ decreases below the torus bifurcation, the rapid spiking begins to modulate in amplitude as the trajectory winds around the attracting torus created near SNp. Further decrease of $b_1$ causes the torus to grow, and eventually parts of the torus shadow, in alternation, the attracting and repelling branches of periodic orbits of the fast system. As $b_1$ decreases further, this leads first to torus canards without heads, then torus canards with heads.

\begin{figure}[t!]
\begin{center}
 \resizebox{6.2in}{!}{\includegraphics{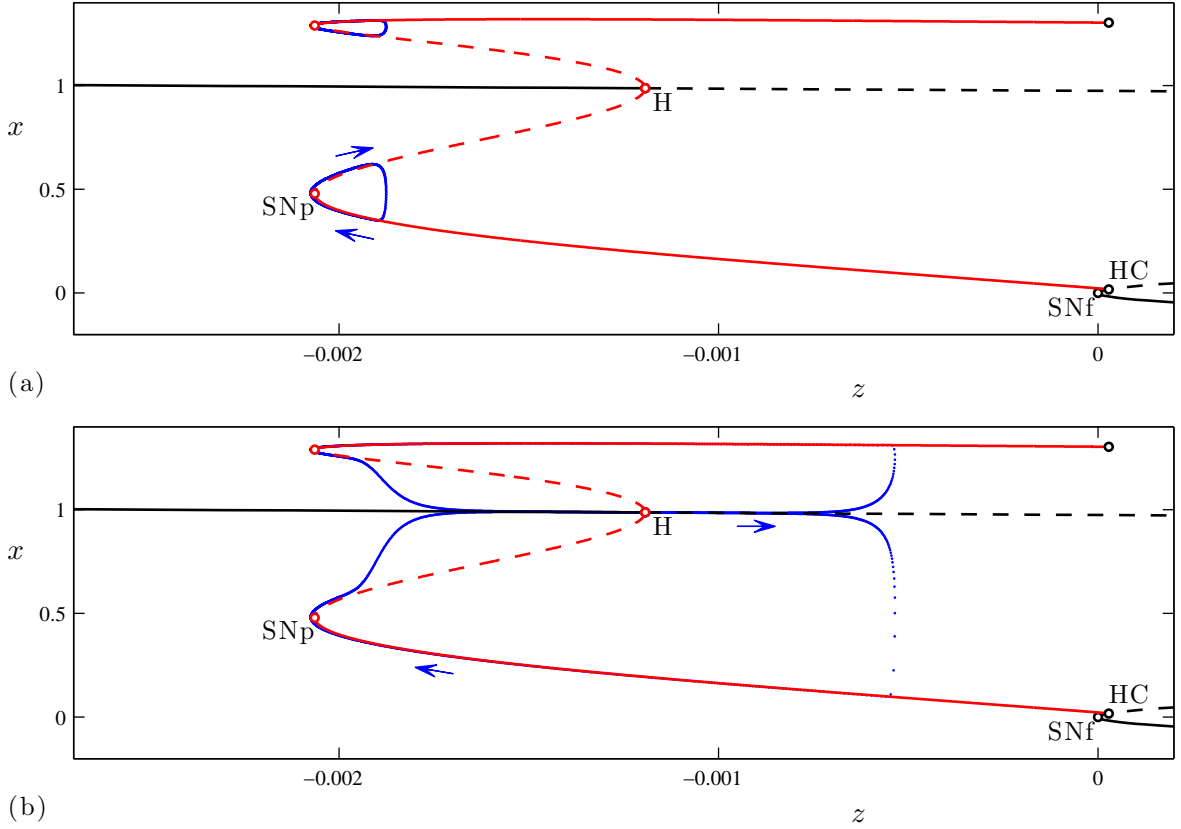}}
\end{center}
\caption{Poincar\'{e} map of torus canard trajectories in the HR system~(\ref{eq:HR}) at $s=-1.95$: (a) torus canard without head at $b_1=-0.16046985$, and (b) torus canard with head at $b_1=-0.16047$. The bifurcation diagrams of the fast system includes branches of fixed points and periodic orbits, and follow the plotting conventions in Fig.~\ref{fig:Purkinje}. The time series of these torus canard orbits are shown in Fig.~\ref{fig:HR_TCtimeseries}.}
\label{fig:HR_TCbif}
\end{figure}

\begin{figure}[t!]
\begin{center}
 \resizebox{5.5in}{!}{\includegraphics{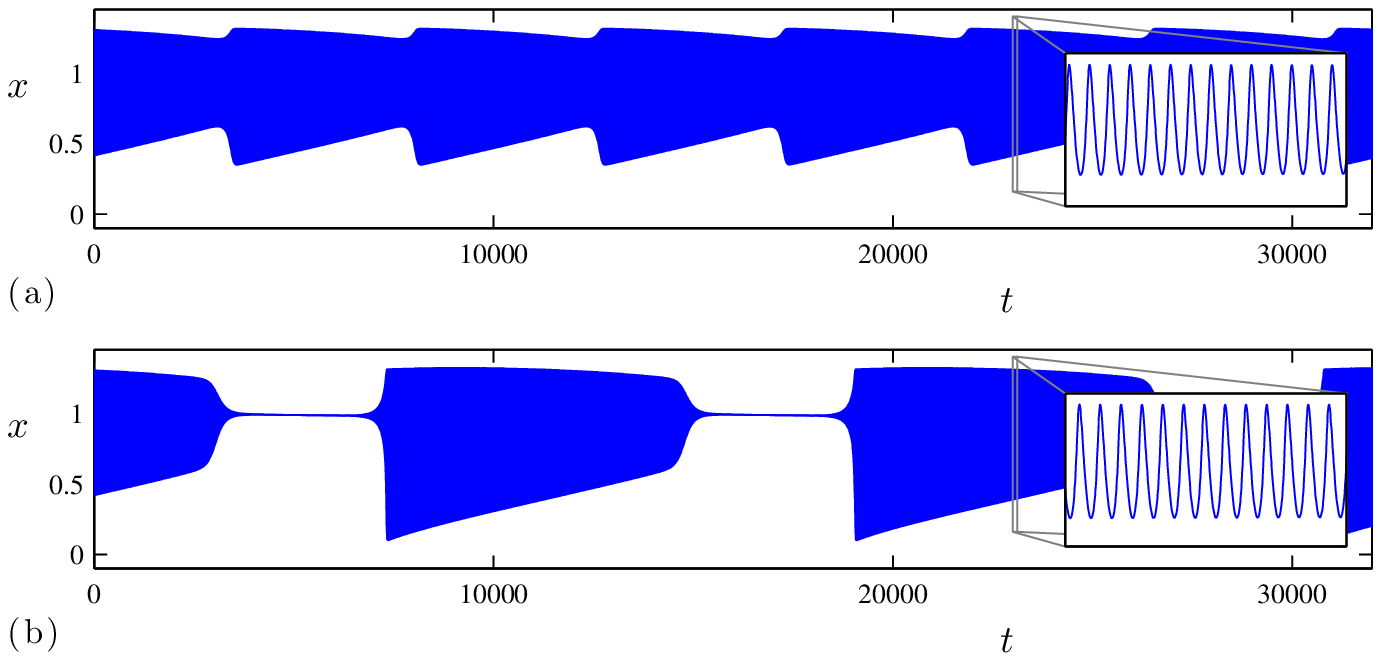}}
\end{center}
\caption{Time series of torus canard orbits in the HR system~(\ref{eq:HR}). Panel (a) shows the headless torus canard from Fig.~\ref{fig:HR_TCbif}a in which the slow modulation in amplitude of the rapid spiking behavior is apparent. Panel (b) shows the torus canard with head from Fig.~\ref{fig:HR_TCbif}b, which qualitatively resembles the bursting solutions found at lower $b_1$ values.}
\label{fig:HR_TCtimeseries}
\end{figure}

To illustrate these dynamics, Fig.~\ref{fig:HR_TCbif} shows two torus canards in projection onto the $(z,x)$ phase space, and Fig.~\ref{fig:HR_TCtimeseries} shows the corresponding time series for the $x$ coordinate. Both types of torus canards spiral on the fast time scale, following the envelope of the outer (attracting) branch of periodic orbits of the fast system to the fold SNp and then continuing for some time along the envelope of the inner (repelling) branch of periodic orbits. The trajectory shown in Fig.~\ref{fig:HR_TCbif}a at $b_1=-0.16046985$ leaves the branch of repelling periodic orbits and returns directly to the attracting branch of periodic orbits, forming a headless torus canard. As $b_1$ is decreased, the length of time that the headless torus canard orbit spends near the branch of repelling periodic orbits increases. Further decrease in $b_1$ results in a narrow region of MMO behavior (not shown), followed by torus canards with heads, as shown in Fig.~\ref{fig:HR_TCbif}b at $b_1=-0.16047$. Now, the trajectory leaves the branch of repelling periodic orbits for the branch of attracting fixed points. The trajectory then drifts up in $z$, leaves the branch of fixed points after a slow passage through the Hopf bifurcation, returns to the branch of attracting periodic orbits, and the cycle repeats.

This bifurcation sequence, consisting of a family of headless torus canards (Fig.~\ref{fig:HR_TCbif}a) followed by MMO and a family of torus canards with heads (Fig.~\ref{fig:HR_TCbif}b), constitutes a torus canard explosion. Moreover, the torus canard explosion marks the transition regime from spiking to sub-Hopf/fold cycle bursting: when $b_1$ is sufficiently negative (i.e., sufficiently past the torus canard explosion), the trajectory does not follow the branch of repelling periodic orbits and instead falls directly off the saddle-node of periodic orbits, resulting in a large amplitude bursting orbit such as the one shown in Fig.~\ref{fig:HR_SHFCburst} at $b_1=-0.162$.

\subsection{Relation to Other Types of Bursting} \label{sec:HR_other}

\begin{figure}[t!]
\begin{center}
 \resizebox{6.2in}{!}{\includegraphics{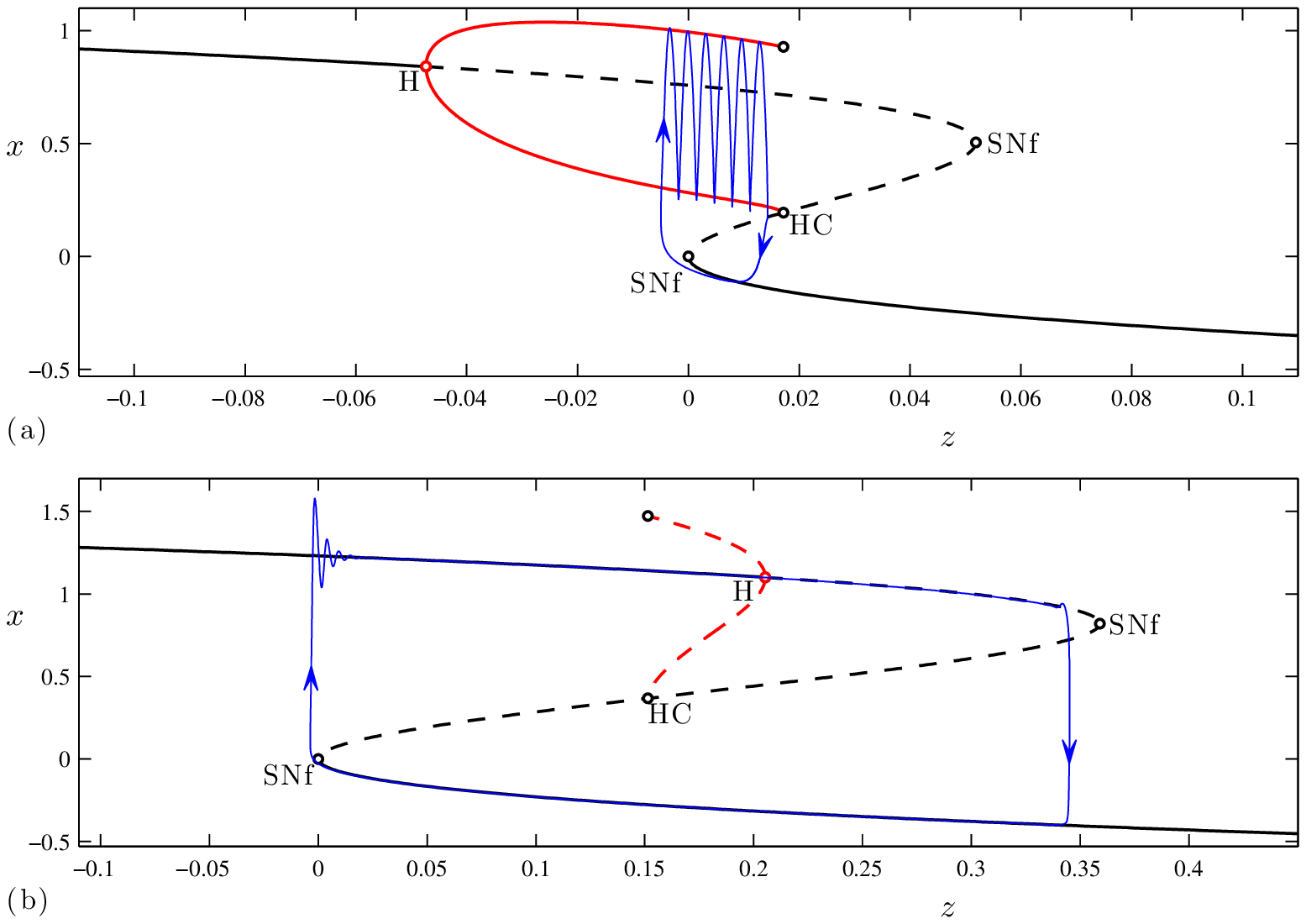}}
\end{center}
\caption{Examples of different bursting types in the HR system~(\ref{eq:HR}): (a) square-wave bursting for $s=-1.61$, and (b) pseudo-plateau bursting for $s=-2.6$. Other parameters are as in Eq.~(\ref{eq:HR_params}), except $b=1$, $\eps = 0.004$, and $b_1=-0.03$. Plotting conventions follow Fig.~\ref{fig:Purkinje}.}
\label{fig:HR_OtherBursters}
\end{figure}

The HR system~(\ref{eq:HR}) exhibits a wide range of different bursting behavior beyond the sub-Hopf/fold cycle bursting described above. Some of this behavior can be understood by considering how varying the parameter $s$ changes the bifurcation structure of the fast system, as in Fig.~\ref{fig:HR_twopar2d}. For example, increasing $s$ eliminates SNp by changing the Hopf bifurcation from subcritical to supercritical. This can lead to the square-wave bursting shown in Fig.~\ref{fig:HR_OtherBursters}a. There, the active phase of the burst is initiated at a saddle-node of fixed points SNf and terminates at a homoclinic bifurcation HC. The classification of this burster is now fold/homoclinic, and an essential ingredient for torus canards --- a saddle-node of periodic orbits in the fast system --- is lost. Therefore, the torus canard phenomenon is also lost. This type of burster has been studied in Refs.~\cite{ML,Shorten,Teka,Tsaneva}.

Decreasing the parameter $s$ also eliminates the saddle-node of periodic orbits. In this case, the Hopf bifurcation H remains subcritical and the saddle-node of periodic orbits SNp is eliminated when it collides with the homoclinic bifurcation HC. This can lead to pseudo-plateau bursting, as shown in Fig.~\ref{fig:HR_OtherBursters}b, which has been studied extensively in Refs.~\cite{Teka,Tsaneva}. In this case, the active phase of the burst again initiates at the saddle-node of fixed points SNf, but these oscillations (which are associated with the complex eigenvalues of the upper fixed point, not the periodic orbits) terminate after the slow passage through the subcritical Hopf bifurcation. Here again, the elimination of an essential ingredient --- the saddle-node of periodic orbits --- results in the loss of the torus canard phenomenon.

In conclusion, the HR system exhibits different types of bursting behavior depending on the choice of parameter $s$. For a wide range of $s$, fold/fold cycle bursting occurs. We showed that, for this type of bursting, a torus bifurcation occurs between the regimes of rapid spiking and bursting, and that a torus canard explosion separates the two.

\section{Torus Canards in the Morris-Lecar-Terman System} \label{sec:MLT}

In this section we consider a version of the Morris-Lecar system~\cite{ML} extended to $\mathbb{R}^3$ by Terman~\cite{Terman}, which we call the Morris-Lecar-Terman (MLT) model. The equations are
\begin{subequations} \label{eq:MLT}
\begin{align}
 \dot{V} &=  y - g_\mathrm{L} (V-E_\mathrm{L}) - g_\mathrm{K} w (V - E_\mathrm{K}) - \gCa m_\infty(V) (V-E_\mathrm{Ca}) \\
 \dot{w} &= - \frac{ w - w_\infty(V)}{\tau_w (V)} \, , \\
 \dot{y} &= \eps (k - V) \, ,
\end{align}
\end{subequations}
where
\begin{subequations} \label{eq:MLT_GatingVars}
\begin{align}
 m_{\infty}(V) &= \frac{1}{2} \left[ 1+\tanh \left( \frac{V-c_1}{c_2} \right) \right] \, , \\
 w_{\infty}(V) &= \frac{1}{2} \left[ 1+\tanh \left( \frac{V-c_3}{c_4} \right) \right] \, , \\
 \tau_w (V)  &= \tau_0 \, \, \sech\!\!\left( \frac{V-c_3}{2 c_4} \right) \, .
\end{align}
\end{subequations}

The MLT model exhibits a wide variety of bursting dynamics. It was examined by Terman~\cite{Terman} in a parameter regime in which it exhibits fold/homoclinic bursting. In addition, the same model was used in Ref.~\cite{Izhikevich} to illustrate both circle/fold cycle bursting and fold/homoclinic bursting. Here, we focus on system~(\ref{eq:MLT}) as an example of the former, in which the active phase of the burst initiates in a saddle-node bifurcation on an invariant circle (i.e., SNIC) and terminates in a fold of limit cycles. We find torus canards in this model, precisely in the transition regime from spiking to this type of bursting.

This section follows the same outline used in the previous section. First, we show that the fast system of~(\ref{eq:MLT}) has a fold of limit cycles (Section~\ref{sec:MLT_fast}) and that the full MLT model has a torus bifurcation (Section~\ref{sec:MLT_tr}). Once these key ingredients are identified, we show that this system includes a torus canard explosion in the transition regime between spiking and bursting (Section~\ref{sec:MLT_tc}).

In what follows, we treat $k$ and $\gCa$ as the primary and secondary control parameters, respectively. The remaining system parameters are fixed at
\begin{subequations} \label{eq:MLT_params}
\begin{gather}
 g_\mathrm{L} = 0.5 \, , \quad g_\mathrm{K} = 2 \, , \quad
 E_\mathrm{L} = -0.5 \, , \quad E_\mathrm{K} = -0.7 \, , \quad E_\mathrm{Ca} = 1 \, , \\
 c_1 = -0.01 \, , \quad c_2 = 0.15 \, , \quad c_3 = 0.1 \, , \quad c_4 = 0.16 \, , \quad \tau_0 = 3 \, , \quad \eps = 0.003 \, ,
\end{gather}
\end{subequations}
which are the values used in Ref.~\cite{Izhikevich}.

\subsection{Bifurcation Analysis of the Fast System} \label{sec:MLT_fast}

\begin{figure}[t!]
 \center
 \resizebox{6.2in}{!}{\includegraphics{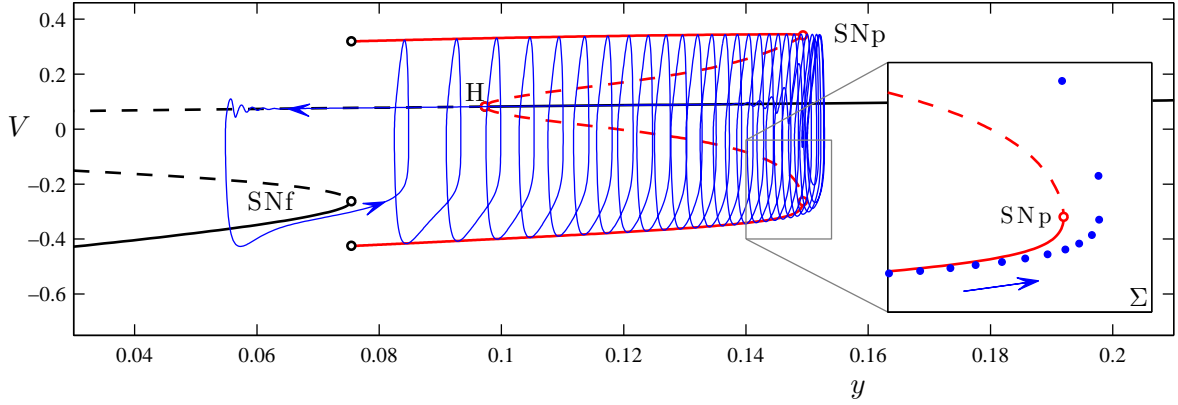}}
  \caption{An example of circle/fold cycle bursting in the MLT system~(\ref{eq:MLT}) at $(k,\gCa)=(-0.0375, 1.25)$. The bursting trajectory (blue curve) is plotted in projection onto the $(y,V)$ phase space along with the bifurcation diagram of the fast system at this value of $\gCa$. Plotting conventions follow Fig.~\ref{fig:Purkinje}. The inset shows the Poincar\'{e} map of the bursting trajectory near SNp, also plotted in projection onto the $(y,V)$ phase space. The Poincar\'{e} surface $\Sigma$ is chosen so that the iterates correspond to local extrema in $V$ of the trajectory.}
  \label{fig:MLT_CircleFoldCycle}
\end{figure}

The fast system of~(\ref{eq:MLT}) is obtained by setting $\eps=0$. The resulting 2D system in $(V,w)$ is the familiar Morris-Lecar system~\cite{ML}. It is independent of $k$, so the slow variable $y$ serves as the bifurcation parameter. A typical bifurcation diagram for the fast system is shown in Fig.~\ref{fig:MLT_CircleFoldCycle}. The branch of fixed points includes two saddle-node bifurcations SNf (with only the right one shown in the figure) and a subcritical Hopf bifurcation H. The branch of periodic orbits created in the Hopf bifurcation undergoes a saddle-node bifurcation SNp then terminates in a SNIC bifurcation as the branch of attracting periodic orbits collides with the
right saddle-node of fixed points. 

Figure~\ref{fig:MLT_CircleFoldCycle} also includes a trajectory of the full MLT system, which illustrates circle/fold cycle bursting. The active phase of the burst ends when the trajectory falls off the branch of attracting periodic orbits at SNp and drifts down in $y$ along a branch of attracting fixed points. The slow passage takes the trajectory through the Hopf bifurcation H and eventually to the lower (stable) branch of fixed points. It then drifts up in $y$ and off the fold of fixed points SNf that is associated with the SNIC, and finally is captured by the attracting branch of periodic orbits, which corresponds to the initiation of the active phase of the burst. Because the active phase of the burst initiates at the SNIC and terminates at the saddle-node of periodic orbits SNp, this is a circle/fold cycle burster in the classification scheme of Ref.~\cite{Izhikevich}.

\begin{figure}[t!]
 \center
 \resizebox{4.5in}{!}{\includegraphics{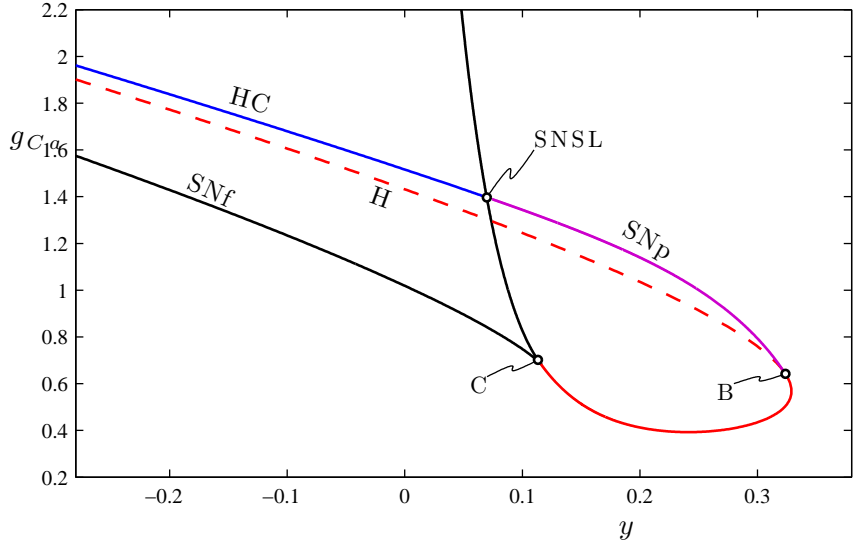}}
  \caption{Two parameter bifurcation diagram of the fast system of~(\ref{eq:MLT}). The loci of saddle-nodes of fixed points SNf merge in a cusp bifurcation C at $(y,\gCa) \simeq (0.1133, 0.7016)$, which creates the locus of Hopf bifurcations H. The saddle-node of periodic orbits SNp emerge from the Bautin bifurcation B at $(y,\gCa) \simeq (0.3238, 0.6418)$ and extend to the saddle-node separatrix loop bifurcation SNSL at $(y,\gCa) \simeq (0.06972, 1.397)$, which also creates a locus of homoclinic bifurcations HC. There is a SNIC bifurcation on the segment of SNf between C and SNSL. Plotting conventions follow Fig.~\ref{fig:HR_twopar2d}.}
  \label{fig:MLT_twopar2d}
\end{figure}

Figure~\ref{fig:MLT_twopar2d} shows how the various codimension-1 bifurcations from Fig.~\ref{fig:MLT_CircleFoldCycle} change as the secondary bifurcation parameter $\gCa$ varies. At large $\gCa$, the saddle-node of periodic orbits SNp disappears when it collides with the saddle-node of fixed points SNf associated with the SNIC in a codimension-2 bifurcation which is of saddle-node separatrix loop type, similar to what is studied in Ref.~\cite{Schecter}; see also Ref.~\cite{HS}. Above this value of $\gCa$, the branch of periodic orbits terminates in a homoclinic bifurcation HC involving the saddle fixed point. At smaller $\gCa$, the two saddle-nodes of fixed points collide in a cusp bifurcation C, which generates a second, supercritical Hopf bifurcation. Below the cusp, the SNIC is no longer possible and the branch of periodic orbits terminates instead in the newly formed Hopf bifurcation. There is also a codimension-2 Bautin bifurcation B as the original Hopf bifurcation changes from subcritical to supercritical, and the saddle-node of periodic orbits SNp terminates at this point. Thus, the MLT system includes a saddle-node of periodic orbits over a wide range of $\gCa$ values ($0.6418 \leq \gCa \leq 1.397$), and it is within this range that we expect the system may also include torus canards.

\subsection{Torus Bifurcation in the Full System} \label{sec:MLT_tr}

\begin{figure}[t!]
 \center
 \resizebox{3.2in}{!}{\includegraphics{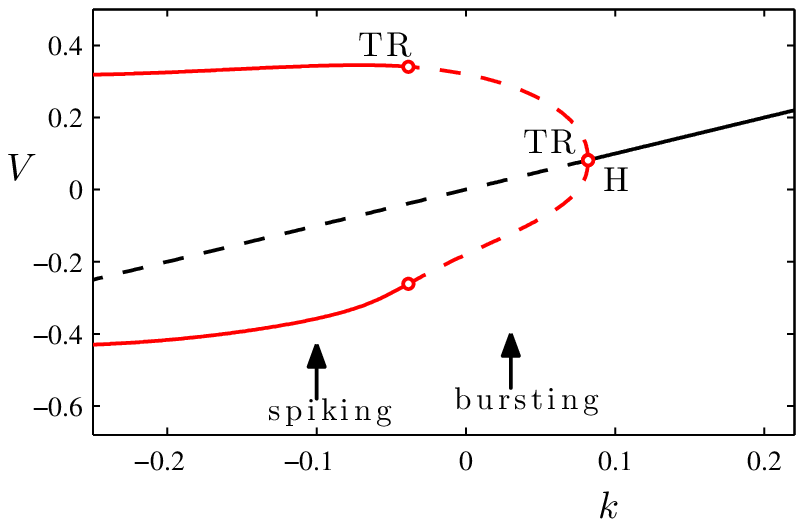}}
  \caption{Bifurcation diagram of the MLT system~(\ref{eq:MLT}) at $\gCa=1.25$, including branches of fixed points (black curve) and periodic orbits (two red curves, indicating maximal and minimal values of $V$ over the orbit). Solid/dashed curves indicate stable/unstable solutions. The torus bifurcation at $k \simeq -0.03852$ is supercritical, leading to bursting at larger $k$.}
  \label{fig:MLT_bif3d}
\end{figure}

The bifurcation diagram of the full MLT model~(\ref{eq:MLT}) is shown in Fig.~\ref{fig:MLT_bif3d} at fixed $\gCa=1.25$, with $k$ as the bifurcation parameter. 
For sufficiently negative values of $k$, the system includes an unstable fixed point and a stable, large amplitude periodic orbit that corresponds to the rapid spiking state of the neuronal system. As $k$ increases, the periodic orbit
%
loses stability in a torus
bifurcation. This is the torus bifurcation we will focus on in this
section. Beyond this torus bifurcation value, the
periodic orbits restabilize in a second torus bifurcation. Finally, for
slightly larger $k$, just beyond this second torus bifurcation, there is a
Hopf bifurcation. The periodic orbits disappear in this Hopf
bifurcation, and the fixed points become stable. This highly depolarized (i.e., large $V$) fixed point corresponds to the physiological state of depolarization block in the MLT system.

The bifurcation sequence presented in Fig.~\ref{fig:MLT_bif3d} persists for a range of $\gCa$ values surrounding $\gCa = 1.25$, but varying this secondary parameter does lead to new behavior.
At smaller values of $\gCa$ the two torus bifurcations coalesce (at $\gCa \simeq 0.67$), and eventually even the Hopf bifurcation disappears. At larger values of $\gCa$ the crucial torus bifurcation disappears in a codimension-2 bifurcation (at $\gCa \simeq 1.30$) that creates a pair of period-doubling bifurcations on the branch of periodic orbits. Thus, the torus bifurcation of interest persists over the range $ 0.67 < \gCa < 1.30$. The behavior beyond this upper limit, in the period-doubling regime, is outside the scope of this paper.

\subsection{Torus Canard Explosion} \label{sec:MLT_tc}

The transition near the torus bifurcation at $k \simeq -0.03852$ in Fig.~\ref{fig:MLT_bif3d} from rapid spiking to bursting occurs by way of a torus canard explosion. For values of $k$ below the torus bifurcation, the periodic orbit of the full system (i.e., the rapid spiking state) is stable. As $k$ increases above the torus bifurcation, the system exhibits amplitude modulated spiking as the trajectory winds around the torus near the saddle-node of periodic orbits of the fast system. The torus grows as $k$ increases, and parts of the torus shadow the attracting and repelling branches of periodic orbits of the fast system in alternation. Further increases in $k$ lead the system through the torus canard explosion, including first the torus canards without heads, then MMO, torus canards with heads, and finally the complete circle/fold cycle bursters, such as the one shown in Fig.~\ref{fig:MLT_CircleFoldCycle}.
Therefore, the torus canards play a central role in the transition from spiking to circle/fold cycle bursting in this model, just as was the case for the HR model in the transition to sub-Hopf/fold cycle bursting.

\begin{figure}[t!]
 \center
 \resizebox{5.5in}{!}{\includegraphics{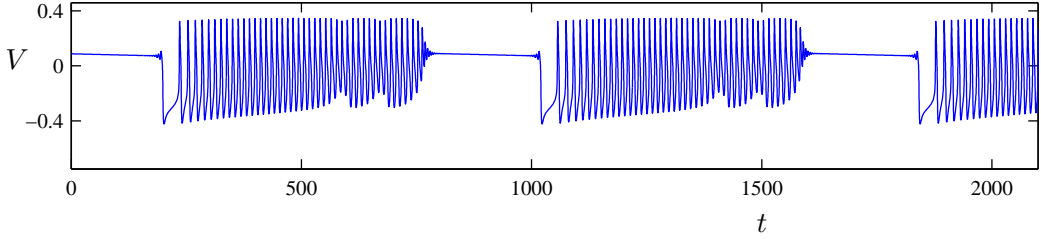}}
  \caption{Time series of $V$ for a MMO in the MLT system~(\ref{eq:MLT}), at $\gCa = 1.25$ and $k=-0.0380$.}
  \label{fig:MLT_MMO}
\end{figure}

Figure~\ref{fig:MLT_MMO} shows the time series of a trajectory at a value of $k$ during the torus canard explosion where the system exhibits MMO dynamics. Each time the trajectory passes through the saddle-node of periodic orbits it transitions from the branch of attracting to the branch of repelling periodic orbits of the fast system, but the direction in which the trajectory leaves the repelling branch of periodic orbits varies from one pass to the next. When it falls outward toward the attracting branch of periodic orbits, it resembles the amplitude modulated spiking and headless torus canard behavior seen at slightly smaller $k$ values. When it falls inward toward the branch of fixed points, the trajectory resembles the bursting and torus-canard-with-head trajectories seen at slightly larger $k$ values. 

\subsection{Relation to Other Types of Bursting} \label{sec:MLT_other}

\begin{figure}[t!]
 \center
 \resizebox{6.2in}{!}{\includegraphics{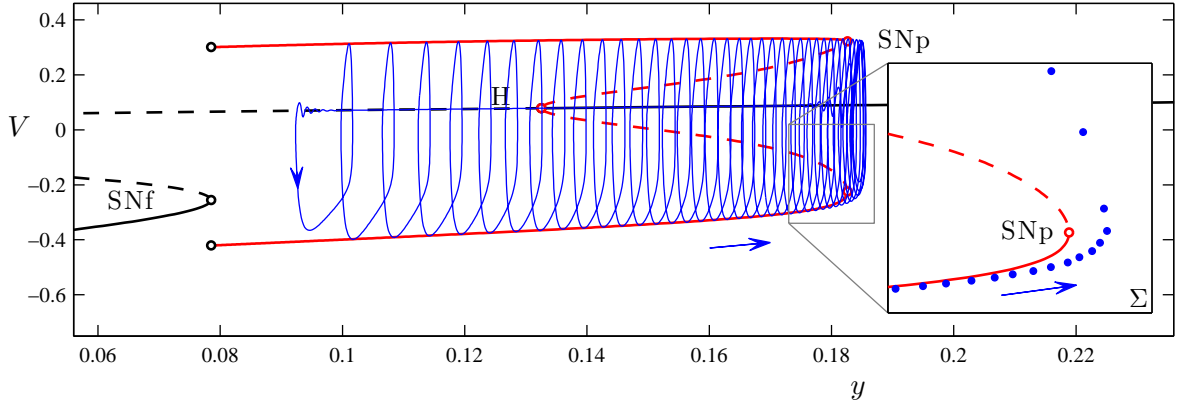}}
  \caption{An example of sub-Hopf/fold cycle bursting in the MLT system~(\ref{eq:MLT}) at $(k,\gCa)=(-0.01, 1.18)$. The bursting trajectory (blue curve) is plotted in projection onto the $(y,V)$ phase space, along with the bifurcation diagram of the fast system at this value of $\gCa$. Plotting conventions follow Fig.~\ref{fig:Purkinje}.}
  \label{fig:MLT_SubHopfFoldCycle}
\end{figure}

In addition to the circle/fold cycle bursting described above, 
the MLT system also exhibits sub-Hopf/fold cycle bursting similar to that observed in the HR model in Section~\ref{sec:HR}. An example of sub-Hopf/fold cycle bursting in the MLT system is shown in Fig.~\ref{fig:MLT_SubHopfFoldCycle}. At this value of $\gCa$, the Hopf bifurcation H is located farther in $y$ from the saddle-node of fixed points SNf that is associated with the SNIC (compare to Fig.~\ref{fig:MLT_CircleFoldCycle}) so the slow passage through the Hopf bifurcation does not take the trajectory to sufficiently small $y$ to involve the SNIC. Instead, the bursting initiates when the trajectory spirals away from the unstable fixed point directly toward the attracting branch of periodic orbits of the fast system. We note however that the transition (as the parameter $k$ increases) from spiking to the sub-Hopf/fold cycle bursting in Fig.~\ref{fig:MLT_SubHopfFoldCycle} goes by way of a torus canards explosion, just as it did for the circle/fold cycle bursting in Fig.~\ref{fig:MLT_CircleFoldCycle}. In both cases, the torus canard explosion is associated with the dynamics near SNp.

In summary, the MLT system exhibits different types of bursting behavior depending on $\gCa$. There is a wide range of $\gCa$ values for which the system exhibits some type of bursting involving a fold of limit cycles --- either circle/fold cycle bursting or sub-Hopf/fold cycle bursting. In each case, the regimes of rapid spiking and bursting are separated by a torus canard explosion.

{\bf Remark: } The details of the transition from rapid spiking to rest as $k$ decreases (in the neighborhood of the lower Hopf bifurcation in Fig.~\ref{fig:MLT_bif3d}) are beyond the scope of this paper, but may involve another bursting regime, as in Ref.~\cite{Terman}.

\section{Torus Canards in the Wilson-Cowan-Izhikevich System} \label{sec:WCI}

In this section, we consider the following extended version of the Wilson-Cowan model~\cite{WilsonCowan} proposed by Izhikevich in Ref.~\cite{Izhikevich}, which we call the Wilson-Cowan-Izhikevich (WCI) system:
\begin{subequations} \label{eq:WCI}
\begin{align}
\dot{x} & =  -x + S(r_x + ax - by + u) \, ,\\
\dot{y} & =  -y + S(r_y + cx - dy + f u) \, ,\\
\dot{u} & = \eps(k - x) \, , 
\end{align}
\end{subequations}
where $S(x) = 1/(1+\exp(-x))$. With $\eps \ll 1$ the variables $x$ and $y$ are fast and $u$ is slow.

As with the models considered in the previous sections, the WCI model can exhibit a wide variety of bursting dynamics. We are interested in this model as an example of a fold/fold cycle burster, where the active phase of the burst initiates in a fold of fixed points and terminates in a fold of limit cycles.

The analysis of this model follows the same steps used in the previous sections. We first show that the fast system of~(\ref{eq:WCI}) has a fold of limit cycles (Section~\ref{sec:WCI_fast}) and that the full system includes a torus bifurcation (Section~\ref{sec:WCI_tr}), then describe the associated torus canards (Section~\ref{sec:WCI_tc}) that exist in the transition from spiking to fold/fold cycle bursting.

We treat $k$ and $r_x$ as the primary and secondary control parameters, respectively, and fix 
\begin{gather} \label{eq:WCI_par}
 r_y=-9.7 \, , \quad a=10.5 \, , \quad b=10 \, , \quad c=10 \, , \quad d=-2 \, , \quad f=0.3 \, , \quad \eps=0.03 \,,
\end{gather}
for the remaining parameters.

\subsection{Bifurcation Analysis of the Fast System} \label{sec:WCI_fast}

\begin{figure}[t!]
\begin{center}
 \resizebox{6.2in}{!}{\includegraphics{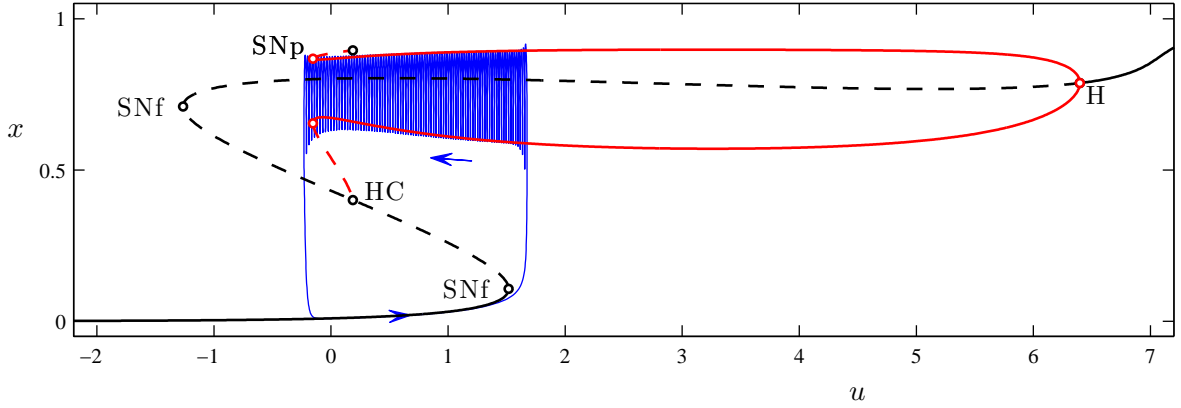}}
\end{center}
\caption{An example of fold/fold cycle bursting in the WCI model~(\ref{eq:WCI}) at $(k,r_x)=(0.6,-4.76)$; other parameters are listed in Eq.~(\ref{eq:WCI_par}). The bursting trajectory (blue curve) is plotted in projection onto the $(u,x)$ phase space, along with the bifurcation diagram of the fast system at this value of $r_x$. Plotting conventions follow Fig.~\ref{fig:Purkinje}.}
\label{fig:WCI_FFCburst}
\end{figure}

The bifurcation diagram of the fast system of~(\ref{eq:WCI}) is shown in Fig.~\ref{fig:WCI_FFCburst}, where the slow variable $u$ serves as the bifurcation parameter. The fixed point is stable at large $u$, but loses stability in a supercritical Hopf bifurcation H as $u$ decreases. The branch of fixed points restabilizes after two saddle-node bifurcations SNf. The branch of periodic orbits created in the Hopf bifurcation undergoes a saddle-node bifurcation SNp and terminates in a homoclinic bifurcation HC involving the saddle fixed point.

Figure~\ref{fig:WCI_FFCburst} also shows a trajectory of the full WCI system which illustrates fold/fold cycle bursting. The active phase of the burst initiates when the trajectory drifts up in $u$ and off the branch of fixed points at a saddle-node of fixed points. During the active phase, the rapid spiking shadows the branch of stable periodic orbits of the fast system, and the slow variable $u$ decreases. The active phase terminates when the trajectory drifts down and off the branch of periodic orbits at SNp, and returns to the stable branch of fixed points to repeat the cycle.

\begin{figure}[t!]
\begin{center}
 \resizebox{4.5in}{!}{\includegraphics{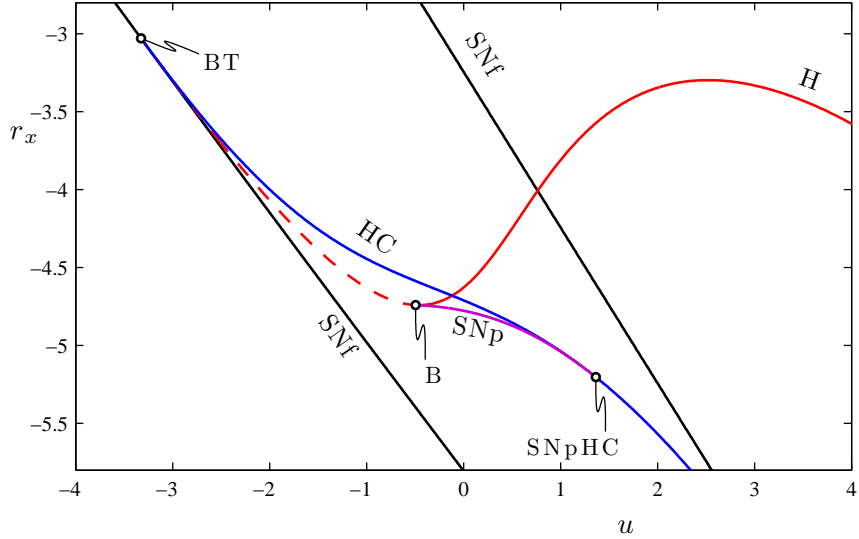}}
\end{center}
\caption{Two-parameter bifurcation diagram of the fast system of~(\ref{eq:WCI}) in the $(u,r_x)$-plane, including loci of saddle-nodes of fixed points SNf, Hopf bifurcations H, saddle-nodes of periodic orbits SNp, and homoclinic bifurcation HC. There are three labeled codimension-2 bifurcations: a Bogdanov-Takens bifurcation BT at $(u, r_x) \simeq (-3.325, -3.029)$, a Bautin bifurcation B at $(u,r_x) \simeq (-0.4945, -4.740)$, and SNpHC at $(u,r_x) \simeq (1.364, -5.203)$. Plotting conventions follow Fig.~\ref{fig:HR_twopar2d}.}
\label{fig:WCI_twopar2d}
\end{figure}

Figure~\ref{fig:WCI_twopar2d} shows how the bifurcation structure of the fast system changes with the secondary control parameter $r_x$. Decreasing $r_x$ from $r_x=-4.76$ causes the saddle-node of periodic orbits SNp to disappear when it collides with the homoclinic bifurcation HC; this occurs at the codimension-2 point labeled SNpHC, at $r_x \simeq -5.203$. Increasing $r_x$ from $r_x=-4.76$ also causes the saddle-node of periodic orbits to disappear, but by a different mechanism. Increasing $r_x$ decreases the amplitude of the periodic orbits near SNp, and at sufficiently large $r_x$ ($r_x \simeq -4.741$), this amplitude shrinks to zero and the branch of periodic orbits collides with the upper branch of fixed points. This creates two new Hopf bifurcations by splitting the branch of periodic orbits into two pieces, one that connects the original Hopf bifurcation to one of the newly-formed Hopf points, and a second that connects the other newly formed Hopf to the homoclinic orbit HC. The latter branch includes SNp, but a codimension-2 Bautin bifurcation eliminates SNp at a slightly larger value of $r_x$ ($r_x \simeq -4.740$). Thus the saddle-node of periodic orbits persists over the range $-5.203 < r_x < -4.740$. Further increase of $r_x$ eliminates one branch of periodic orbits as the supercritical Hopf bifurcations coalesce. There is also a codimension-2 Bogdanov-Takens bifurcation BT in which the subcritical Hopf and the homoclinic HC disappear.

\subsection{Torus Bifurcation in the Full System} \label{sec:WCI_tr}

\begin{figure}[t!]
\begin{center}
 \resizebox{3.2in}{!}{\includegraphics{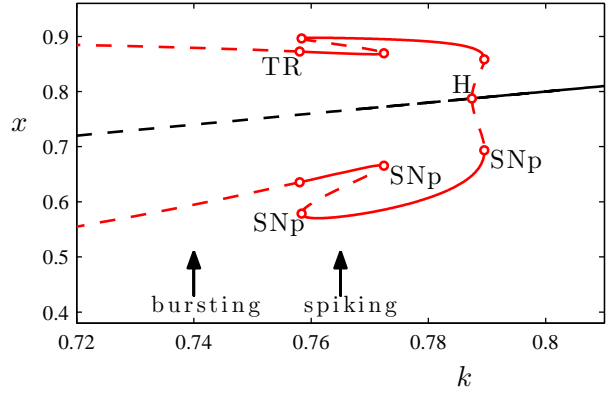}}
\end{center}
\caption{Bifurcation diagram of the WCI system~(\ref{eq:WCI}) at $r_x=-4.76$, upon variations of parameter $k$, including branches of fixed points (black curve) and periodic orbits (two red curves, indicating maximal and minimal values of $x$ over the orbit). Solid/dashed curves indicate stable (unstable) solutions. The torus bifurcation at $k \simeq 0.7580$ is supercritical, leading to bursting at smaller $k$.}
\label{fig:WCI_bif3d}
\end{figure}

The bifurcation diagram of the full WCI model~(\ref{eq:WCI}) is presented in Fig.~\ref{fig:WCI_bif3d} at fixed $r_x=-4.76$, with $k$ as the bifurcation parameter. It shows that this system has a branch of fixed points which loses stability as $k$ decreases in a subcritical Hopf bifurcation H. The branch of periodic orbits that emerges from this Hopf point is unstable at onset, and its stability changes three times in three saddle-node bifurcations. Finally, the branch destabilizes via a torus bifurcation TR at $k \simeq 0.7580$. This torus bifurcation lies between the regimes of spiking and bursting dynamics, and is associated with torus canards.

\subsection{Torus Canard Explosion} \label{sec:WCI_tc}

\begin{figure}[t!]
\begin{center}
 \resizebox{6.2in}{!}{\includegraphics{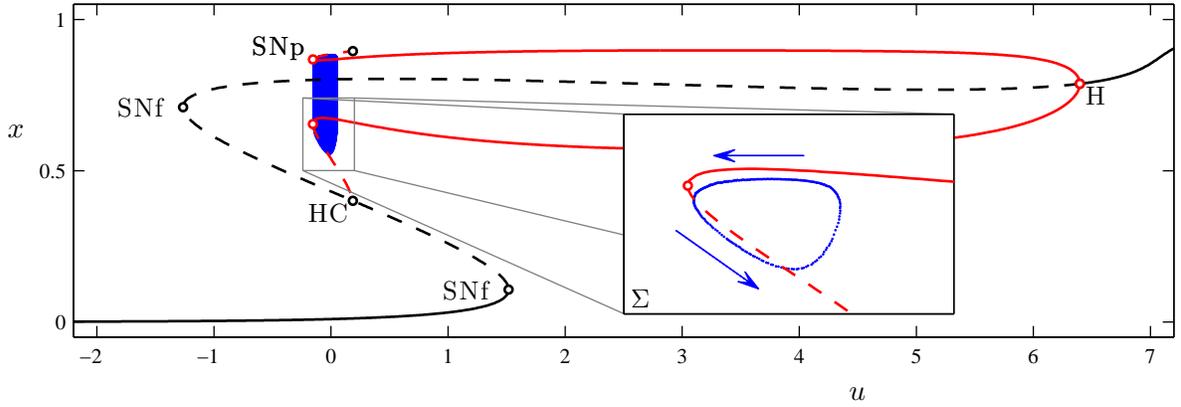}}
\end{center}
\caption{An example of a torus canard orbit (a `duck without head') in the WCI model~(\ref{eq:WCI}) at $(k, r_x ) = (0.7575, -4.76)$. The trajectory is plotted in projection onto the $(u,x)$ phase space, along with the bifurcation diagram of the fast system at this value of $r_x$. Plotting conventions in Fig.~\ref{fig:Purkinje}. The inset shows the Poincar\'{e} map of the torus canard trajectory near SNp, with the Poincar\'{e} surface $\Sigma$ chosen so that the iterates correspond to local extrema in $x$ of the trajectory.}
\label{fig:WCI_toruscanard}
\end{figure}

The transition from rapid spiking to bursting as $k$ decreases through the torus bifurcation in Fig.~\ref{fig:WCI_bif3d} occurs via torus canards. At the torus bifurcation point, the trajectory of the periodic orbit resembles the periodic orbit of the fast system at the saddle-node of periodic orbits of the fast system. At a value of $k$ slightly below the torus bifurcation the trajectory winds around a torus near SNp, spending time, in alternation, near the attracting and repelling branches of periodic orbits of the fast system (see the `duck without head' trajectory shown in Fig.~\ref{fig:WCI_toruscanard}).  Further decrease of $k$ completes the torus canard explosion (including MMO and `duck with head' trajectories, not shown) and leads to the fold/fold cycle bursting trajectory shown in Fig.~\ref{fig:WCI_FFCburst}. Moreover, the behavior at this value of $r_x = -4.76$ is representative of a range of $r_x$ values, and the WCI model~(\ref{eq:WCI}) includes a transition from spiking to fold/fold cycle bursting via a torus canard explosion over a range of $r_x$ values in which the key ingredients for torus canards persist.

\subsection{Relation to Other Types of Bursting}

For values of $r_x$ above the Bautin bifurcation at $r_x = -4.74$, the fast system no longer includes a saddle-node of periodic orbits so bursters involving a `fold cycle' are no longer possible. In this regime, the fast system includes a subcritical Hopf bifurcation (see Fig.~\ref{fig:WCI_twopar2d}), and this can lead to new bursting scenarios. For example, it is possible to have a bursting orbit that follows the branch of attracting fixed points of the fast system down in $u$ to the subcritical Hopf bifurcation and then spirals along the associated branch of repelling periodic orbits for some time.

The saddle-node of periodic orbits SNp persists as $r_x$ decreases down to the SNpHC point. Below this point the active phase of the bursting cycles terminates at the homoclinic orbit (i.e., fold/homoclinic bursting). We note however that the torus bifurcation of the full system only persists down to $r_x \simeq -5.10$. Below this value the stable periodic orbits of the full system lose stability in a period doubling bifurcation instead, so the transition from spiking to bursting does not involve torus canards.

\section{Conclusions} \label{sec:conclusions}

Torus canards were originally identified in a fifth order model of a Purkinje cell~\cite{Kramer}, where it was shown that the torus canard explosion occurs precisely in the transition region between tonic spiking and bursting. Some basic aspects of the dynamics of torus canards were studied in Ref.~\cite{Benes} in the context of an elementary third order model, obtained by rotating a planar bistable system of van der Pol type and introducing symmetry--breaking terms. In this article, we extended this work and presented two primary results. First, we showed that torus canards are common among computational neuronal systems of fast-slow type for which the fast systems have a saddle-node of periodic orbits (a.k.a. a fold of limit cycles) and the full systems have a torus bifurcation. The torus canard orbits spend long times near branches of attracting and repelling periodic orbits of the fast system in alternation, switching over from the former to the latter exactly near the saddle-node of periodic orbits. Moreover, these torus canards are the natural analog in one higher dimension of the by-now classical canards of limit cycle type, which spend long times near branches of attracting and repelling fixed points in alternation, as for example in the van der Pol and FitzHugh-Nagumo equations \cite{Diener,Kakiuchia}. It was shown here that the Hindmarsh-Rose (HR) system, the Morris-Lecar-Terman (MLT) model, and the Wilson-Cowan-Izhikevich (WCI) model all have the essential ingredients to possess torus canards, namely a saddle-node of periodic orbits in the fast system and a torus bifurcation in the full system. Also, we described in detail the families of torus canards that exist in these models, and identified the torus canard explosions.

Second, we demonstrated that the torus canard explosions in these systems play central roles in the transitions between the spiking and bursting regimes. In the HR system, the torus canards occur precisely in the transition region from spiking to sub-Hopf/fold cycle bursting, in which the active phase of the burst initiates when the trajectory passes a subcritical Hopf bifurcation point and terminates when it passes the fold of limit cycles. The transitions from spiking to bursting in the MLT and WCI models are, respectively, to circle/fold cycle bursting in which the active phase initiates in a saddle-node bifurcation on an invariant circle (a.k.a. SNIC), and to fold/fold cycle bursting in which the active phase initiates as the trajectory passes a fold of fixed points.

To conclude this article, we discuss other neuronal systems in which torus canards might occur. First, we think that it is likely that torus canards exit in other models that exhibit the types of bursting --- sub-Hopf/fold cycle, circle/fold cycle, and fold/fold cycle --- that we studied here. For example, the top-hat burster of Best, et al.~\cite{Best} is known to exhibit fold/fold cycle bursting, although there may be some technical differences since this is a fourth-order model. 

Second, there are other classes of bursting dynamics in which the active phase of the burst terminates in a fold of limit cycles, but in which the initiation event is different from those considered here. For example, from the classification in Table 1.6 of Ref.~\cite{Izhikevich}, one sees that there are also super-Hopf/fold cycle bursters. For these, the active phase of the burst initiates with a supercritical Hopf bifurcation. However, since the termination event is also a fold of limit cycles, these bursters should also exhibit torus canards. We note that, for these super-Hopf/fold cycle bursters, the slow passage effect through a Hopf bifurcation will play a role in determining the system parameters for which torus canards exist, just as it did for the sub-Hopf/fold cycle bursters.

Finally, while we have only examined bursters in which the initiation event involves bifurcations of fixed points, there are also bursters in which the burst-phase is triggered by the bifurcation of an invariant set of dimension greater than zero, such as a limit cycle or torus. We think that, as long as the burst phase terminates in a fold of limit cycles, these systems may also exhibit torus canards, as well as new types of canards of mixed type that spend time near other types of attracting and repelling sets, not just limit cycles, and in various sequences.

\vspace{0.2cm}
\noindent
{\bf Acknowledgement}: 
The research of J.B. and A.M.B. was supported by the Center for
BioDynamics at Boston University and the NSF (DMS 0602204, EMSW21-RTG).
The research of M.D was supported by EPSRC under grant EP/E032249/1; 
M.D. is grateful for the hospitality of the Center for BioDynamics 
at Boston University during several visits when part of this work was 
completed. The research of T.K. was
supported by NSF-DMS 1109587. M.A.K. holds a Career Award at the
Scientific Interface from the Burroughs Wellcome Fund.
The authors thank Hinke Osinga and Andrey Shilnikov for useful discussion.

\bibliography{NeuroTC}

\end{document}